\documentclass[preprint,showpacs,showkeys,preprintnumbers,aps]{revtex4}

\usepackage{graphicx}

\begin{document}

\preprint{Balke et al., Co$_2$Mn$_{0.5}$Fe$_{0.5}$Si.}

\title{The half-metallic ferromagnet Co$_2$Mn$_{0.5}$Fe$_{0.5}$Si.}

\author{Benjamin Balke, Hem C. Kandpal, Gerhard H. Fecher,  and Claudia Felser}
\email{felser@uni-mainz.de}
\affiliation{Institut f\"ur Anorganische und Analytische Chemie, \\
Johannes Gutenberg - Universit\"at, D-55099 Mainz, Germany.}

\date{\today}

\begin{abstract}

Electronic structure calculation were used to predict a new material for spintronic applications.
Co$_2$Mn$_{0.5}$Fe$_{0.5}$Si is one example which is stable against on-site correlation and disorder effects due to the position of the Fermi energy in the middle of the minority band gap. Experimentally the sample were made exhibiting $L2_1$ structure and a high magnetic order.
\end{abstract}

\pacs{75.30.-m, 71.20.Be, 61.18.Fs}

\keywords{half-metallic ferromagnets, electronic structure,
          magnetic properties, Heusler compounds, spintronic}

\maketitle

%%%%%%%%%%%%%%%%%%%%%%%%%%%%%%%%%%%%%%%%%%%%%%%%%%%%%%%%%%%%%%%%%%%%

\section{Introduction}
\label{IN}

Half-metallic ferromagnets have been proposed as ideal candidates for spin injection
devices because they have been predicted to exhibit 100~{\%} spin polarization at the 
Fermi energy ($\epsilon_F$) \cite{GME83}. From the applications point of view, a high 
Curie temperature for a half-metallic ferromagnet may be an important condition. 
For this reason, Heusler alloys ($L2_1$ structure) have recently attracted great interest. 
Some of these alloys exhibit high Curie temperatures and, according to theory, should have 
a high spin polarization at the Fermi energy \cite{FKW06}. 
Calculations also show that anti-site disorder will destroy the high spin polarization \cite{PCF04a}
, implying that precise control of the atomic structure of the Heusler alloys is required.

The Heusler alloy Co$_2$MnSi has attracted particular interest because it is predicted to have a large minority spin band gap of 0.4~eV and, at 985~K, has one of the highest Curie temperature,  among the known Heusler compounds \cite{FSI90,BNW00}. Structural and magnetic properties of Co$_2$MnSi have been reported for films and single crystals \cite{RRH02,KHM03,WPK05a}. From tunnelling magneto resistance (TMR) data with one electrode consisting of a Co$_2$MnSi film Sakuraba {\it et al} \cite{SMO06} measured a TMR ration of 159~{\%} at 2K and $\approx70$~{\%} at 300K. 
If using Co$_2$FeSi as one electrode Inomata {\it et al} \cite{IOM06} obtained TMR ratios of 60~{\%} at 5K and 41~{\%} at 300K.

An important quantity for the application of the half-metallic ferromagnets is the size of the gap in the minority states and the position of $\epsilon_F$ inside of the gap. Small gaps may be easily destroyed by temperature effects or quasi-particle excitations \cite{CAK06}. The half-metallicity may also be easily destroyed if $\epsilon_F$ is located close to the band edges, either of the minority valence or conduction bands.

The recent TMR results \cite{SMO06,IOM06} show that elctrodes made of Heusler compounds result in high TMR ratios at low temperatures  but the temperature dependence is still a challenge which has to be solved to use Heusler electrodes in applications.
The present investigation focuses on searching for a mixed compound where the half-metallic behaviour is stable against temperature effects.

The self-consistent electronic structure calculations were carried out using the full potential linearized augmented plane wave method (FLAPW) as provided by Wien2k \cite{BSM01}. The LDA$+U$ method \cite{AAL97} was used to account 
for on-site correlation at the transition metal sites. 
Semi-empirical values corresponding to 7.5\% of atomic values 
of Coulomb-exchange parameter have been used as suggested in our previous publication \cite{KFF06}.

%%%%%%%%%%%%%%%%%%%%%%%%%%%%%%%%%%%%%%%%%%%%%%%%%%%%%%%%%%%%%%%%%%%%

\section{Results and discussion}

\subsection{Electronic properties}

\begin{figure}[t]
\includegraphics[scale =1.0]{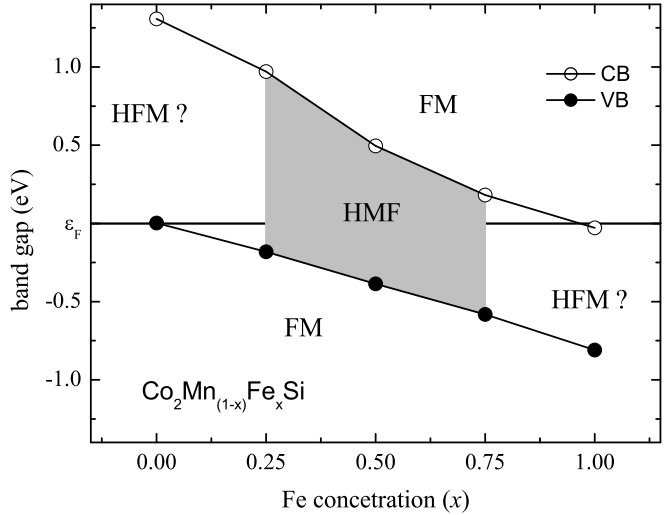}
\centering
\caption{Dependence of the minority band gap on the Fe concentration \textit{x} in Co$_2$Mn$_{1-x}$Fe$_x$Si.
					The extremal energies of the gap involving states are shown. The shaded areas indicate the region 
					of half-metallic ferromagnetism. Lines are drawn for clarity.}
  	\label{fig-1}
  \end{figure}

The size of the minority band gap and position of $\epsilon_F$ can be seen from the calculated energies displayed in Fig.~\ref{fig-1} for the Heusler alloy Co$_2$Mn$_{1-x}$Fe$_x$Si. 
In Co$_2$MnSi, $\epsilon_F$ is close to the top of the valence band. In Co$_2$FeSi, the situation is different and $\epsilon_F$ is near to the bottom of the conduction band. Both of the compounds are at the two corners of the gap. 
In Co$_2$Mn$_{0.5}$Fe$_{0.5}$Si $\epsilon_F$ is directly in the middle of the minority band gap and therefore this compound should be more stable against temperature effects or quasi-particle excitations compared to pure Co$_2$MnSi and Co$_2$FeSi.

%%%%%%%%%%%%%%%%%%%% Figure 1 %%%%%%%%%%%%%%%%%%%%%%%%%%%%%%%%%%%%%%%%%%%%%%%%
%%%%%%%%%%%%%%%%%%%%%%%%%%%%%%%%%%%%%%%%%%%%%%%%%%%%%%%%%%%%%%%%%%%%%%%%%%%%%%

\subsection{Structural and magnetic properties}
Co$_2$Mn$_{0.5}$Fe$_{0.5}$Si samples were prepared by arc melting of stoichiometric amounts of the constituents in an argon atmosphere at 10$^{-4}$~mbar. The polycrystalline ingots were then annealed in an evacuated quartz tube at 1273~K for 21~days. 
The samples exhibiting the Heusler type $L2_1$ structure with a lattice parameter of 5.64~{\AA} (see Fig. 2(a)). Another proof of the high order of the samples are the results from the $^{57}$Fe M{\"o\ss}bauer spectra taken of the samples (see Fig. 2(b)). 
The spectrum is dominated by an intense sextet with a line width of approximately $(0.14\pm0.01)$~mm/s. In addition to the sextet, a much weaker line at the centre of the spectrum is visible. Its contribution to the overall intensity of the spectrum is approximately 3.5~{\%}. The origin of the singlet may be caused by anti-site disorder leading to a small fraction of paramagnetic Fe atoms. The hyperfine field (HFF) at the Fe sites amounts to $26.5\times10^6$~A/m.
For more details and other properties of the whole series of Co$_2$Mn$_{1-x}$Fe$_x$Si with the Fe concentration ranging from $x=0$ to 1 in steps of 0.1 see Ref.\cite{BFK06}.

%%%%%%%%%%%%%%%%%%%% Figure 2 %%%%%%%%%%%%%%%%%%%%%%%%%%%%%%%%%%%%%%%%%%%%%%%%
  \begin{figure}[t]
\includegraphics[scale = 1.0]{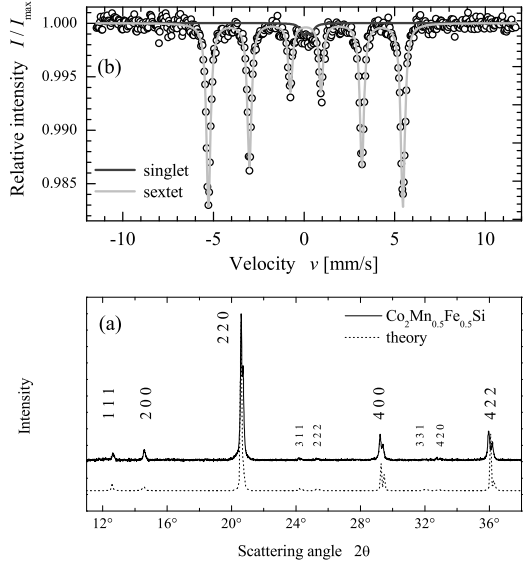}
\centering
\caption{(a) XRD spectra for Co$_2$Mn$_{0.5}$Fe$_{0.5}$Si.
             The spectra were excited by Mo K$_\alpha$ radiation.
				(b) $^{57}$Fe M{\"o\ss}bauer spectrum of Co$_2$Mn$_{0.5}$Fe$_{0.5}$Si.
            The spectrum was taken at 290K and excited by a $^{57}$Co(Rh) source.
            Solid lines are results of a fit to determine the sextet and singlet
            contributions and to evaluate the hyperfine field.}
  	\label{fig-2}
  \end{figure}
%%%%%%%%%%%%%%%%%%%%%%%%%%%%%%%%%%%%%%%%%%%%%%%%%%%%%%%%%%%%%%%%%%%%%%%%%%%%%%

\section{Summary}
Electronic structure calculations predicted the Heusler compound Co$_2$Mn$_{0.5}$Fe$_{0.5}$Si due to the position of $\epsilon_F$ in the middle of the minority band gap as very stable against any kind of effects destroying the halfmetalicity and therefore well suited for spintronic applications.
Polycristaline samples were prepared exhibiting $L2_1$ structure and a high magnetic order.

%in which & is replaced by \& including TEX command. So, when you see URL, 
%please use the following address
%http://authors.elsevier.com/JournalDetail.html?PubID=505704&Precis=AIND


This work is financially supported by the DFG (project TP1 and TP7 in
research group FG 559).


\begin{thebibliography}{0}
\expandafter\ifx\csname natexlab\endcsname\relax\def\natexlab#1{#1}\fi
\expandafter\ifx\csname bibnamefont\endcsname\relax
  \def\bibnamefont#1{#1}\fi
\expandafter\ifx\csname bibfnamefont\endcsname\relax
  \def\bibfnamefont#1{#1}\fi
\expandafter\ifx\csname citenamefont\endcsname\relax
  \def\citenamefont#1{#1}\fi
\expandafter\ifx\csname url\endcsname\relax
  \def\url#1{\texttt{#1}}\fi
\expandafter\ifx\csname urlprefix\endcsname\relax\def\urlprefix{URL }\fi
\providecommand{\bibinfo}[2]{#2}
\providecommand{\eprint}[2][]{\url{#2}}

\end{thebibliography}


\begin{thebibliography}{00}

\bibitem{GME83}
R.A. de Groot, F.M. M\"uller, P.G. van Engen, and K.H.J. Buschow,
{\it Phys. Rev. Lett} {\bf 50} (1983), p. 2024.

\bibitem{FKW06}
G. H. Fecher, H. C. Kandpal, S. Wurmehl, C. Felser, and G. Sch\"onhense,
{\it J. Appl. Phys.} {\bf 99} (2006), 08J106.

\bibitem{PCF04a}
S. Picozzi, A. Continenza, and A. J. Freeman,
{\it Phys. Rev. B.} {\bf 69} (2004), 094423.

\bibitem{FSI90}
S. Fujii, S. Sugimura, S. Ishida, and S. Asano,
{\it J. Phys.: Condens. Matter} {\bf 2} (1990), p. 8583.

\bibitem{BNW00}
P. J. Brown, K.-U. Neumann, P. J. Webster, and K. R. A. Ziebeck, 
{\it J. Phys.: Condens. Matter} {\bf 12} (2000), p. 1827.

\bibitem{RRH02}
M. P. Raphael, B. Ravel, Q. Huang, M. A. Willard, S. F. Cheng, B. N. Das, R. M. Stroud, K. M. Bussmann, J. H. Claassen, and V. G. Harris,
{\it Phys. Rev. B.} {\bf 66} (2002), 104429.

\bibitem{KHM03}
S. K\"ammerer, S. Heitmann, D. Meyners, D. Sudfeld, A. Thomas, A. H\"utten, and G. Reiss,
{\it J. Appl. Phys.} {\bf 93} (2003), p. 7945.

\bibitem{WPK05a}
W. H. Wang, M. Przybylski, W. Kuch, L. I. Chelaru, J. Wang, Y. F. Lu, J. Barthel, H. L. Meyerheim, and J. Kirschner,
{\it Phys. Rev. B.} {\bf 71} (2005), 144416.

\bibitem{SMO06}
Y. Sakuraba, T. Miyakoshi, M. Oogane, H. Kubota, Y. Ando, A. Sakuma, and T. Miyazaki,
{\it Appl. Phys. Lett.} submitted.

\bibitem{IOM06}
K. Inomata, S. Okamura, A. Miyazaki, M. Kikuchi, N. Tezuka, M. Wojcik, and E. Jedryka,
{\it Journal of Physics D: Applied Physics} {\bf 39} (2006), p.816.

\bibitem{CAK06}
L. Chioncel, E. Arrigoni, M. I. Katsnelson and A. I. Lichtenstein, 
{\it Phys. Rev. Lett.} {\bf 96} (2006), 137203.

\bibitem{BSM01}
P. Blaha, K. Schwarz, G. K. H. Madsen, D. Kvasnicka and J. Luitz, 
WIEN2k, An Augmented Plane Wave + Local Orbitals Program for Calculating Crystal Properties, (2001).

\bibitem{AAL97}
V. I. Anisimov, F. Aryasetiawan and A. I. Lichtenstein,
{\it J. Phys. Condens. Matter} 9 (1997), p. 767.

\bibitem{KFF06}
H. C. Kandpal, G. H. Fecher, C. Felser and G. Sch\"onhense,
{\it Phys. Rev. B} {\bf 73} (2006), 094422.

\bibitem{BFK06}
B. Balke, H. C. Kandpal, G. H. Fecher, and C. Felser,
{\it Phys. Rev. B.} submitted, cond-mat/0606108.






\end{thebibliography}
\end{document}